# Magnetization dynamics driven by displacement currents across a magnetic tunnel junction

C. K. Safeer[1], Paul S. Keatley[2], Witold Skowroński[3], Jakub Mojsiejuk[3], Kay Yakushiji[4], Akio Fukushima[4], Shinji Yuasa[4], Daniel Bedau[5], Fèlix Casanova[6,7], Luis E. Hueso[6,7], Robert J. Hicken[2], Daniele Pinna[1], Gerrit van der Laan[8], Thorsten Hesjedal[1]

[1]Clarendon Laboratory, Department of Physics, University of Oxford, Oxford, OX1 3PU, United Kingdom

[2]Department of Physics and Astronomy, University of Exeter, Stocker Road, Exeter, EX4 4QL, United Kingdom

[3]AGH University of Krakow, Institute of Electronics, Al. Mickiewicza 30, 30-059 Kraków, Poland

[4]National Institute of Advanced Industrial Science and Technology, Research Center for Emerging Computing Technologies, Tsukuba, Ibaraki 305-8568, Japan

[5]Western Digital, San Jose Research Center, 5601 Great Oaks Parkway San Jose, CA 95119

[6]CIC nanoGUNE BRTA, 20018 Donostia-San Sebastián, Basque Country, Spain

[7]IKERBASQUE, Basque Foundation for Science, 48009 Bilbao, Basque Country, Spain

[8]Diamond Light Source, Harwell Science and Innovation Campus, Didcot, Oxfordshire OX11 0DE, United Kingdom

## Abstract

Understanding the high-frequency transport characteristics of magnetic tunnel junctions (MTJs) is crucial for the development of fast-operating spintronics memories and radio frequency devices. Here, we present the study of frequency-dependent capacitive current effect in CoFeB/MgO-based MTJs and its influence on magnetization dynamics using time-resolved magneto-optical Kerr effect technique. In our device operating at gigahertz frequencies, we find a large displacement current of the order of mA's, which does not break the tunnel barrier of the MTJ. Importantly, this current generates an Oersted field and spin-orbit torque, inducing magnetization dynamics. Our discovery holds promise for building robust MTJ devices operating under high current conditions, also highlighting the significance of capacitive impedance in high frequency magnetotransport techniques.

**Introduction**

Magnetic tunnel junctions (MTJs) serve as fundamental components in various spintronics devices, mainly relying on the tunneling magnetoresistance (TMR) effect [1–4]. In TMR devices, the electrical resistance depends on the relative alignment of the moments in the magnetic layers, facilitating efficient data read-out in memory devices such as hard disk drives [5] and magnetic random-access memory (MRAM) [6]. Moreover, the spin torque effects generated by current [7] or voltage [8] in MTJs can be used to electrically switch the magnetization. The speed of both reading and writing processes plays a pivotal role in the development of fast-operating memories, as needed in modern devices operating at gigahertz (GHz) frequencies.

In addition to data storage and memory applications, MTJs are also building blocks for dynamic radio frequency (RF) spintronic components. Primarily, these include spin torque nano-oscillators [9–11] that exhibit autonomous precession of the magnetization by leveraging anti-damping spin-transfer torques, energy harvesters [12], RF detectors [13,14] and magnonic devices that utilize the propagation and manipulation of spin waves in ferromagnetic materials [15]. Regarding MTJ nano-oscillators, their capacity to detect and manipulate RF signals at the nanoscale holds promise for applications in RF signal processing, wireless communication [16], and emerging neuromorphic technologies [17]. Magnonic devices are envisioned to be pivotal for the development of low-power electronics in the future [15]. In addition to these practical applications, experimental RF magnetotransport techniques, such as spin-torque ferromagnetic resonance (ST-FMR) [18,19], have been extensively employed to gain insight into fundamental spintronics principles, particularly current-induced magnetization dynamics in various systems. All these factors underscore the practical and fundamental significance of studying frequency-dependent physical effects in MTJs.

A well-known frequency-dependent effect observed in MTJs is the magnetocapacitance effect [20,21], where MTJs can behave like leaky capacitors. Analogously to TMR, they exhibit the tunnel magnetocapacitance (TMC) effect [20,22], where the overall capacitance depends on the relative alignment of the magnetization (parallel or antiparallel) of the free and fixed magnetic layers within the MTJ. The strength of the TMC is strongly dependent on frequency [22], and has even been reported to reverse sign within a specific frequency range [23,24]. Importantly, the capacitive reactance of MTJs varies with frequency, with low reactance expected at high frequencies [20,22–24] leading to significant displacement currents. Displacement current is not an actual charge flow across a capacitor; instead, it was introduced according to the Ampère-Maxwell equation to explain the magnetic fields resulting from changing electric fields in a dielectric medium, such as the MgO tunnel barrier in our MTJ device. The large displacement current in MTJ circuits at high frequencies can be harnessed to induce magnetization dynamics. To the best of our knowledge, this observation has not been previously reported.

In this article, using specially designed CoFeB/MgO/CoFeB-based MTJ devices, we report an unprecedented study of the magnetization dynamics induced by displacement currents across an MTJ. Our findings reveal that substantial displacement currents, in the range of mA's, flow through the MTJ circuit at GHz frequencies. Remarkably, this substantial current does not damage the MTJs, as the displacement current does not constitute charge passing through the dielectric of the capacitor (i.e., the MgO barrier). This discovery offers great potential for the development of robust MTJ devices capable of operating under high current conditions. Additionally, our research underscores the significance of considering capacitive impedance when analyzing high frequency magnetotransport data.

**Device fabrication and TMR measurements**

Figures 1(a) and 1(b) show an optical image and a cross-section schematic diagram of the MTJ device used in this study, respectively. First, a multilayer of W (5 nm)/CoFeB (1.3 nm)/MgO (2.5 nm)/CoFeB (5 nm)/Ta (5 nm)/Ru (5 nm) was deposited on top of an oxidized Si substrate using magnetron sputtering. Then, using a two-step lithography and etching process, the bottom W/CoFeB/MgO heterostructure was patterned into a microwire, while the following top CoFeB/Ta/Ru layers were patterned into a nanopillar [Fig. 1(b)]. Subsequently, 20 nm $SiO_2$ was deposited to electrically isolate a 40-nm-thick Pt top electrode from the bottom W/CoFeB electrode. In this study, we used a device with an MTJ diameter of ~1.5 μm and a resistance-area product of the 2.5-nm-thick MgO tunnel barrier of 1 MΩ · μm$^2$.

We then performed DC magnetotransport measurements to determine the variation of the TMR as a function of the in-plane magnetic field ($B_y$). Figure 1(c) shows the measurements of the MTJ device used in this experiment. In the bottom W/CoFeB/MgO multilayer, the CoFeB thickness is 1.3 nm, for which the total magnetic anisotropy is expected to be a mixture of bulk in-plane and interfacial out-of-plane anisotropy [25]. The bulk in-plane anisotropy component is almost uniform along different in-plane directions in the $x - y$ plane. The device was then annealed at 400°C in a high-vacuum furnace to improve both TMR and interfacial out-of-plane anisotropy [26] (see Sec. I of the Supplemental Material). The top CoFeB layer is 5 nm thick and is therefore dominated by bulk in-plane anisotropy. The combination of anisotropies of the bottom and top CoFeB layers leads to the observed TMR curve as a function of $B_y$ shown in Fig. 1(c), where the remanent in-plane magnetization of the top CoFeB layer forms an angle between 0° to 90° with the out-of-plane tilted remanent magnetization of the bottom CoFeB layer, giving a high resistance state. The bottom layer can be easily saturated into a parallel in-plane magnetization direction by applying approximately

>|20| mT, giving low resistance states. A TMR ratio up to 70% was observed between the remanent and saturated magnetic states. The large TMR confirms an efficient spin tunneling through MgO while the variation of the TMR with respect to the magnetic field confirms the remanent magnetization states of the two CoFeB layers and their variation under the applied magnetic field. The net out-of-plane tilted anisotropy of the bottom CoFeB layer is ideal to obtain magnetization dynamics under voltage and current applications, which will be discussed in the later sections.

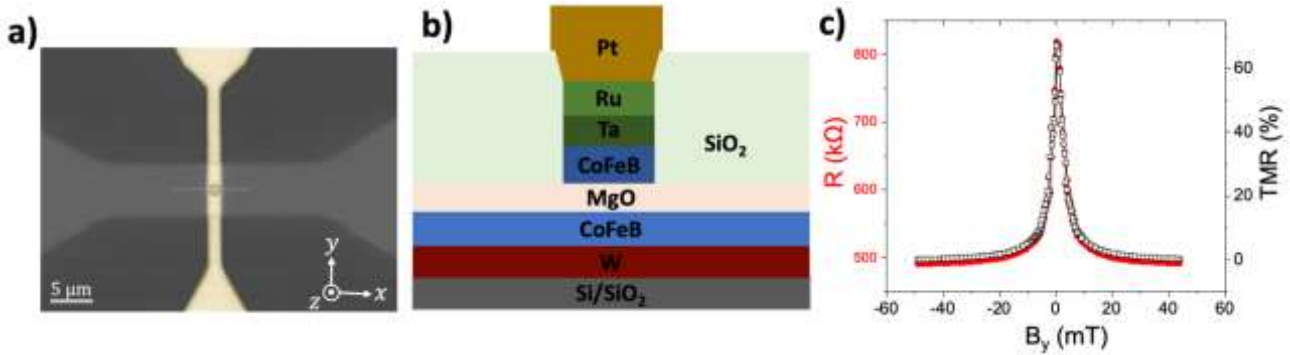

*Figure 1: (a) Top view (optical micrograph) of the 1.5 μm-diameter device with the Pt top electrode in the center highlighted in yellow and the MTJ pillar in faint green, respectively. The bottom W/CoFeB/MgO microwire is perpendicular to the Pt electrode. (b) Schematic diagram of the cross-section of the layer structure at the device region marked by white dotted line in panel (a). (c) Magnetotransport measurements showing variation of resistance (red, left axis) and corresponding TMR (black, right axis) as a function of in-plane magnetic field.*

**Study of magnetization dynamics**

The magnetization dynamics in the bottom W/CoFeB microwire region was explored under different electrical excitations using time-resolved magneto-optical Kerr effect (TR-MOKE) microscopy. Magnetization dynamics were probed using ultrafast laser pulses derived from a mode locked fiber laser operating at a wavelength of 1040 nm and repetition rate of 80 MHz. The laser pulses with nominal duration of 140 fs were frequency doubled by means of optical second harmonic generation to produce probe pulses with wavelength of 520 nm. The beam was passed along a quad-pass optical delay line for up to 8 ns of optical time delay with sub-ps temporal resolution. The beam was expanded ×2 to reduce beam divergence, linearly polarized, and focused to a diffraction limited spot using a long working distance ×50 (numerical aperture 0.55) microscope objective lens to achieve a spatial resolution of ~400 nm. The reflected probe laser pulses were collected using the same lens and the change in polarization detected using a polarizing balanced photodiode bridge detector. Measurements were made in the polar MOKE configuration that allows the out-of-plane component of the dynamic magnetization ($\Delta M_z$) [27] to be detected when the device is magnetized in-plane by an in-plane field generated by a calibrated permanent magnet assembly. The TR-MOKE microscope was used for two types of measurements [20], i.e., to reveal either the temporal or spatial evolution of magnetization dynamics by fixing either the position or the time delay, respectively. In the first case, the probe laser spot was fixed at a selected position on the bottom electrode, as shown in Fig. 2(a), and the variation in the TR-MOKE signal as a function of time delay was recorded. In the second case, the sample position was scanned beneath the probe laser spot while the time delay remained fixed, allowing the spatial character of magnetization dynamics at a particular time delay to be imaged across the entire bottom electrode.

Magnetization dynamics were excited by applying either an electrical voltage impulse or a microwave voltage waveform to the device. Voltage impulses synchronized to the laser repetition rate at 80 MHz were produced with nominal rise time, duration, and amplitude of 30 ps, 70 ps, and -5 V respectively. The resulting broadband excitation stimulates precession of the bottom electrode magnetization at the ferromagnetic resonance (FMR) frequency corresponding to the bias magnetic field applied to the device. To isolate the spatial character of the FMR mode, a microwave comb generator was used to apply a single frequency microwave voltage waveform to the device. A microwave comb containing multiples of 80 MHz from 160 MHz to 18 GHz was generated from an 80 MHz input signal derived from the laser repetition rate. The microwave comb was filtered using narrow pass notch filters with <80 MHz bandwidth to leave just the frequency component corresponding to the FMR. The amplitude of the microwave waveform was adjusted using a combination of microwave amplification and attenuation. For each excitation type the voltage impulse or waveform was passed through a directional coupler for monitoring using a sampling oscilloscope. A bias tee was used to decouple the impulse and comb generators from the device, to minimize the risk of electrostatic damage to the device when making electrical connections, while allowing the TMR to be monitored using the DC port of the bias tee. Finally, the output of the impulse or comb generator was amplitude modulated at ~31.4 kHz, leading to a modulation of the device excitation and resulting polar Kerr signal, which was then recovered using a lock-in amplifier.

The unique geometry of our device, as illustrated in Figs. 1(b) and 2(a), offers two distinct advantages. Firstly, it enables the application of electrical input either vertically through the MTJ from the top to the bottom electrode, or horizontally through the top or bottom electrodes separately. Secondly, since the bottom CoFeB electrode extends beyond the width of the MTJ and non-magnetic top contact, it is possible to investigate magnetization dynamics in the vicinity of the MTJ, namely the propagation of spin waves that may originate from beneath the pillar and propagate along the bottom CoFeB electrode, and any steady-state magnetization dynamics in the entire bottom CoFeB region [28]. Considering these possibilities, we investigated magnetization dynamics within the microwire region of the W/CoFeB by applying two types of electrical input signals: (i) the voltage impulse of 30 ps rise time and 70 ps duration; and (ii) an RF excitation of fixed frequency (2 GHz). The results of each of these cases are discussed separately below.

**Magnetization dynamics under a voltage pulse**

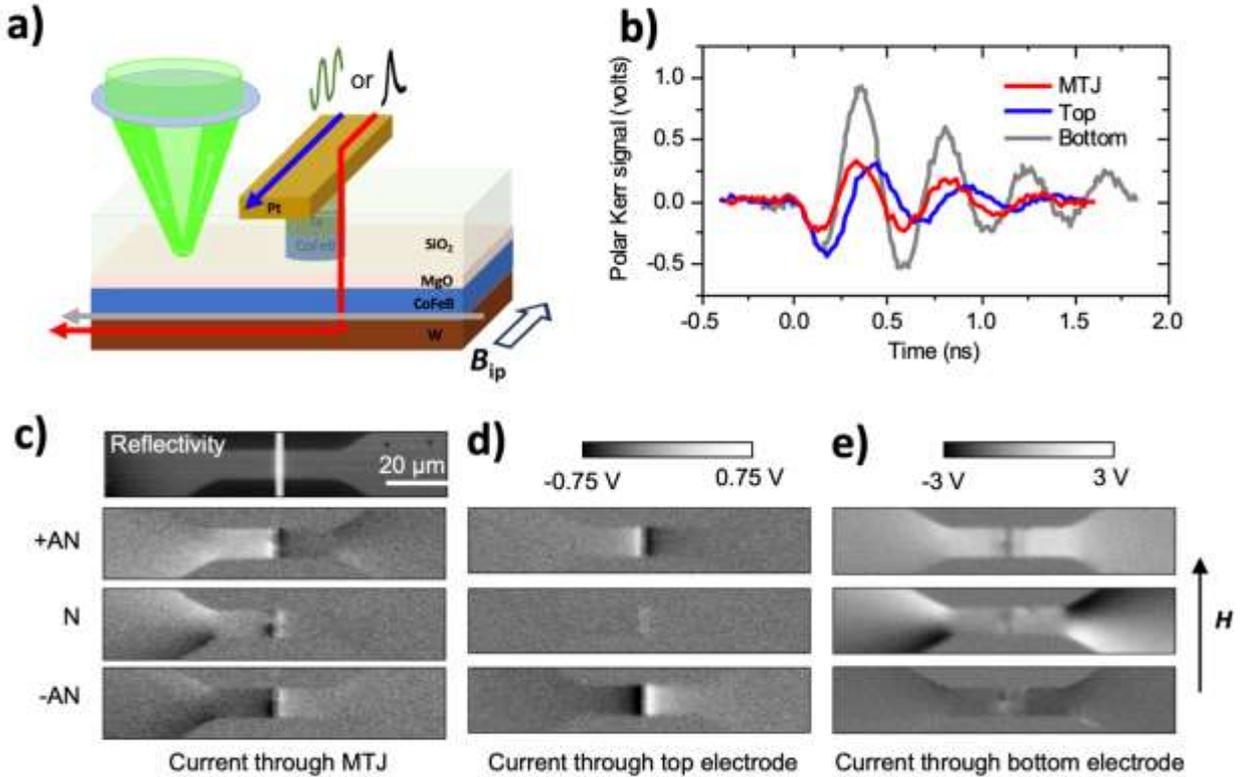

*Figure 2: (a)* Schematic view of the TR-MOKE experiment in which the input is either applied in the form of a voltage pulse (black) or RF wave (green). The red, blue, and grey arrows represent the current paths for the input signals applied through the MTJ, top and bottom electrode, respectively. For the first case (red), the current path also includes part of the top and bottom electrodes. *(b)* Plot of the Kerr rotation vs. time delay taken after fixing the laser spot at the left side of the top electrode as shown in Fig. 1(b) and applying an input voltage pulse across the MTJ (red), top (blue) and bottom electrode (grey). Spatially resolved TR-MOKE images taken at a fixed time delay correspond to the first +antinode (+AN), node (N), and -antinode (-AN) of the plot shown in b, after applying the input voltage across *(c)* the MTJ, *(d)* the top and *(e)* bottom electrode, respectively. The optical reflectivity image of the device is shown in the top panel of (c), where the bottom CoFeB microwire electrode (grey with tapered section to left and right of center), on which the TR-MOKE scanning was performed, and the position of the top electrode (orthogonal white line at the center)

In the presence of $B_y = 75$ mT, a 70 ps voltage pulse with an amplitude of 930 mV was applied across the MTJ, with the expected current path indicated by the red arrow in Fig. 2(a). Subsequently, we measured the corresponding TR-MOKE variation on the left side of the MTJ, ~1 μm away from the top electrode [Fig. 2(a)]. As shown by the red curve in Fig. 2(b), we observed an oscillating TR-MOKE signal that gradually decayed over time, confirming the detection of magnetization precession. The time required to complete one cycle of precession is ~0.5 ns, corresponding to a frequency of approximately 2 GHz. Subsequently, we confirmed that this frequency matched the resonance frequency at $B_y = 75$ mT (see Sec. II of the Supplemental Material). This means that while the broadband voltage pulse can, in principle, induce magnetization precession at random multiple frequencies, only the precessional component at 2 GHz that is at resonance for the applied field of $B_y$ = 75 mT contributes to the observed TR-MOKE signal.

To understand the spatial distribution of the magnetization precession, we then conducted TR-MOKE scans of the entire W/CoFeB microwire region as shown in Fig. 2(c), at three different time delays which correspond to nodes and antinodes of the time-varying signal in Fig. 2(b). Interestingly, we observed a predominant spin precession in the immediate vicinity of the top electrode (vertical strip at the center of the reflectivity image), with opposite $\Delta M_z$ on its left and right sides. Any magnetization dynamics beneath the top contact were not observed since the Pt top contact thickness was larger than the optical skin depth resulting in a narrow vertical strip of negligible signal in the TR-MOKE images.

To identify the mechanism giving rise to the results shown in Fig. 2(c), we first need to consider four different scenarios through which the applied voltage pulse can induce magnetization dynamics, as illustrated in Fig. 3(a-d): (i) voltage-controlled magnetic anisotropy (VCMA) variation at the CoFeB/MgO interface underneath the pillar region, launching propagating spin waves [29]; (ii) spin-transfer torque (STT) in the pillar region inducing propagating spin waves [30]; (iii) spin-orbit torque (SOT) inducing magnetization precession in the entire W/CoFeB microwire region [31] (see Sec. IV of the Supplemental Material for details); and (iv) an Oersted field from the top electrode inducing spin precession in the W/CoFeB microwire region. We have performed micromagnetic simulations (shown in the Supplemental Material, section VIII) to further verify the expected magnetization dynamics depicted in Figs. 3(a-d). Among the above four possibilities, we can conclusively determine that the observation in Fig. 2(c) is related to case (iv). For cases (i) and (ii), we would observe spin wave propagation toward both sides of the top electrode with the same sign for $\Delta M_z$. For case (iii), we should observe steady-state spin precession with the same sign of $\Delta M_z$ along the entire W/CoFeB microwire region in which the current flows (from the left side to the top electrode). Thus, the only effect that can result in spin precession with opposite $\Delta M_z$ on the two sides of the top electrode, as observed in Fig. 3(a), is related to case (iv). This is because the Oersted field induced by the current on the left and right sides of the top electrode is pointing in opposite directions. However, to generate a large enough Oersted field to create such magnetization dynamics, a high current of the order of mA's is typically required (see Sec. III of the Supplemental Material). Considering that the MTJ resistance is ~490 kΩ for $B_y$ > 20 mT, as determined in the DC-TMR measurement [Fig. 1(c)], the maximum current expected from a voltage pulse of 930 mV is ~1.86 μA. This current is too small to generate a strong enough Oersted field, suggesting that a much larger current must flow through the top electrode when applying a high-frequency pulse.

To verify the above scenario, we repeated the same TR-MOKE measurement protocol with the voltage pulse being applied solely through the top electrode [indicated by the blue arrow in Fig. 2(a)], which has a resistance of ~110 Ω. We then adjusted the amplitude of this voltage pulse to ~375 mV to achieve a similar TR-MOKE signal amplitude to that observed when applying 930 mV across the MTJ, as shown in Fig. 2(b). This corresponds to a current flow of ~3.5 mA, which unambiguously confirms that when applying a 930-mV-pulse across the MTJ, the actual current flowing through the top electrode is more than three orders of magnitude larger than the expected current extracted from DC measurements (1.86 μA), indicating the presence of a frequency-dependent impedance effect in our MTJ circuit.

When a voltage pulse is applied across the MTJ, the current also flows through both the top and bottom electrodes as well, as indicated by the red arrow in Fig. 2(a). In addition to the TR-MOKE contrast observed due to the Oersted field-induced magnetization dynamics near the top electrode, TR-MOKE contrast was also observed in the bottom electrode (W/CoFeB) region, as shown in Fig. 2(c). The MOKE contrast is more pronounced on the left-hand side of the top electrode where a large current flows. The bottom electrode consists of three regions: a straight wire region in the middle where the current is expected to flow along the $x$-axis, and two triangular-shaped regions on either side where the current flow has both $x$ and $y$ components. Two important features were observed in the TR-MOKE images shown in Figure 2(c): (i) a large contrast where the current diverges from the middle straight wire region to the triangular regions; and (ii) opposite contrasts along the two diagonals of the triangular region. Both features can be indistinguishably attributed to either Oersted field or SOT-induced magnetization dynamics (for details, see Sec. IV and VIII of the Supplemental Material). If an Oersted field is generated due to current flowing through the bottom electrode, its $y$-component acts on the magnetization in the middle wire region, while both $x$ and $y$-components act on magnetization in the triangular region. However, since magnetization is fixed along $y$ by the external applied field, only the $x$-component of the Oersted field creates magnetization dynamics along $z$, resulting in the observed MOKE contrast being larger in the triangular region. The $x$-component of the Oersted field is opposite along the two diagonals in the same triangular region, resulting in the observed opposite contrasts. Interestingly, the action of damping-like SOT creating an effective out-of-plane field and thus magnetization dynamics along $+z/-z$ directions, has the same symmetry as the Oersted field-induced magnetization dynamics (for details, see Sec. IV of the Supplemental Material). Therefore, it is not possible to distinguish between the two effects. However, as for the case of the top electrode explained earlier, only a current on the order of mA's generates the large Oersted field required for magnetization dynamics. Similarly, SOT-induced magnetization dynamics in W/CoFeB usually require a large current density on the order of at least $10^{10}$ A/m$^2$ [32], and considering the dimensions of our device, a current on the order of mA's will be required. This further confirms a large current flow of the order of mA in our MTJ circuits.

To further confirm the origin of the magnetization dynamics of the bottom electrode, we repeated the TR-MOKE experiment by applying a voltage pulse (corresponds to a current of ~1 mA) only across the bottom electrode [indicated by grey arrow in Fig. 2(a)] and the resulting TR-MOKE image is shown in Fig.2 (e). Again, a large TR-MOKE contrast was observed in a triangular region as well as opposite contrasts along the two diagonals of the triangular region. This further confirms the action of either Oersted field or/and SOT-induced magnetization dynamics in the bottom electrode (see Sec. IV of the Supplemental Material).

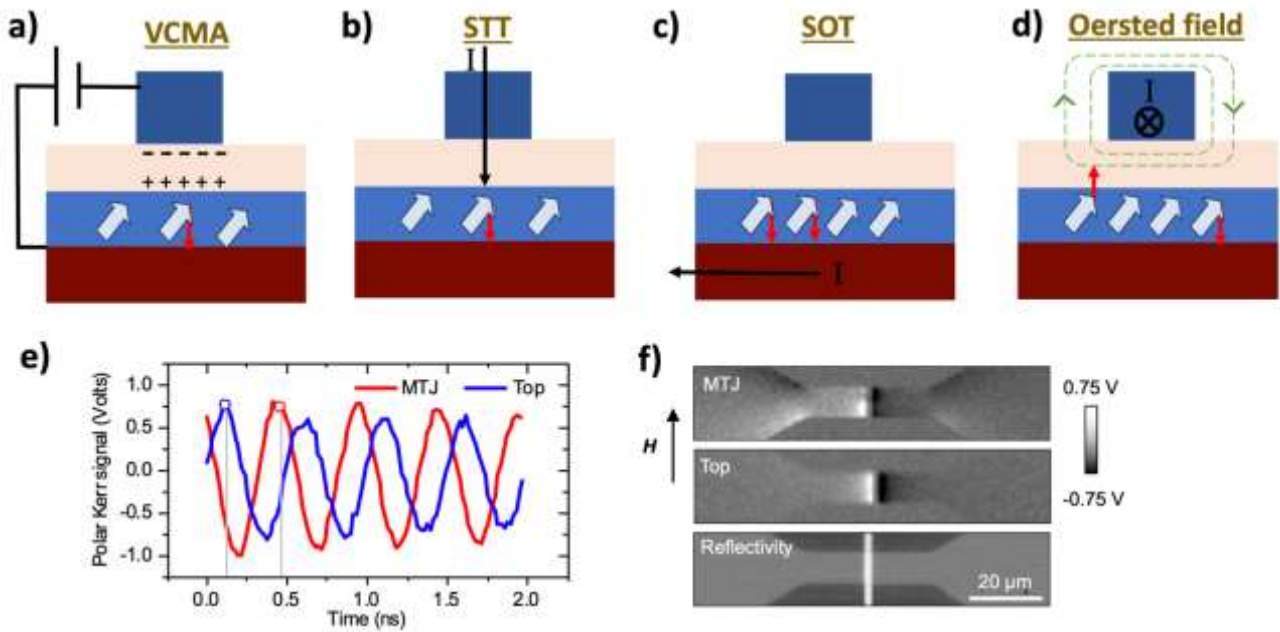

*Figure 3: Schematic diagram illustrating the possible origins of magnetization dynamics (white arrows) under and on the two sides of the top electrode: (a) VCMA, (b) STT, (c) SOT, and (d) Oersted field. (e) Polar MOKE signal vs. time delay when applying a 2 GHz input voltage across the MTJ (red) and the top electrode (blue), respectively. (f) Corresponding spatial scans for the same input voltage applied across the MTJ (above) and the top electrode (below), respectively, taken at a fixed time delay indicated in panel (e).*

**Magnetization dynamics for RF wave input**

As shown above, the application of a 70 ps voltage pulse initiates magnetization dynamics at the resonance frequency of 2 GHz. However, given that a voltage pulse should excite a broad band of frequencies, it is possible that only a fraction of the voltage amplitude (930 mV across the MTJ and 375 mV across the top electrode) contributes to the magnetization dynamics at 2 GHz, resulting in a smaller TR-MOKE signal (per input voltage). To validate the full effect of the amplitude of the high-frequency input signal, we conducted the same experiments shown in Fig. 2, but this time using a sine wave input signal at 2 GHz [indicated by the green wave in Fig. 2(a)]. As expected, the magnetooptical signal/input voltage ratio is larger in this case [Fig. 3(e)]. A continuously oscillating

TR-MOKE signal without decay was obtained, confirming the one-to-one correlation between the continuous precession of magnetization and the continuous input wave signal. Importantly, similar to the observation in Fig. 2, the TR-MOKE spatial scan [Fig. 3(f)] displays opposite precession signals near the two sides of the top electrode, confirming the Oersted field-induced magnetization dynamics. The amplitudes of the TR-MOKE signals, when applying the input wave with a peak-to-peak voltage ($V_{p-p}$) of ~550 mV across the MTJ (corresponding to $I \approx 1.1$ µA) and 250 mV across the top electrode (corresponding to $I \approx 2.2$ mA), were almost the same, as shown by the red and blue curves in Fig. 3(e), respectively. This again confirms that at high frequencies, the total impedance in our circuit is more than three orders of magnitude lower than in the DC resistance case.

**Discussion**

To elucidate our findings, it is necessary to examine an MTJ circuit model in which the overall impedance diminishes as the frequency increases. One plausible approach involves accounting for the capacitance effect of the MTJ. In this context, we can regard the MTJ as a parallel resistance-capacitance (RC) circuit, with MgO functioning as the dielectric of the capacitor. The expression for the overall magnitude of the impedance of the MTJ is:

$$Z = \frac{1}{\sqrt{(\frac{1}{R})^2 + (2\pi f C)^2}} \quad ,$$

where $f$ and $C$ are frequency and capacitance, respectively. For a parallel plate capacitor with one plate having an infinite area, the geometrical capacitance ($C$) can be written as (for details see Sec. V of the Supplemental Material):

$$C = \frac{2\varepsilon_0 \varepsilon_r A}{d} \quad ,$$

where $A$ and $d$ are the area of MTJ pillar and MgO thickness, respectively. Thus, $Z$ depends on the frequency and area that is proportional to the square of the radius of the device, which is ~750 nm for our MTJ. For $V=$ 930 mV, we calculated the overall current flow at different frequencies ($I=V/Z$) as shown in Fig. 4 for different MTJ sizes with the green curve representing our MTJ. We clearly see that in the GHz regime (indicated by grey area in Fig. 4), the current flow in the circuit is on the order of a few mA, instead of the µA range expected from the DC measurements. In our device, the substrate Si/SiO$_2$ can also act as a capacitor, allowing displacement current across them in parallel to the MTJ circuit, depending on the resistivity of Si. This may influence the total displacement current in the MTJ circuit, as explained in the supplemental materials section IX.

The current flowing through the capacitor at high frequencies, i.e., the displacement current, does not involve the actual flow of charges through the dielectric barrier. The RC time constant defines the charging and discharging of the capacitor (see Sec. VI of the Supplemental Material), which is ~55 ns for our device. This time constant is significantly larger than the period of the input wave, which is 0.5 ns at 2 GHz. At such a short timescale, the accumulation of charge responsible for the resistance in the capacitor (MgO) is expected to be extremely low. This, in turn, permits a significant displacement current flow in the outer circuit of the MTJ capacitor. This outer circuit includes the top and bottom electrodes, and it is this large displacement current that gives rise to the observed magnetization dynamics, attributed to the induced Oersted field and SOT effects.

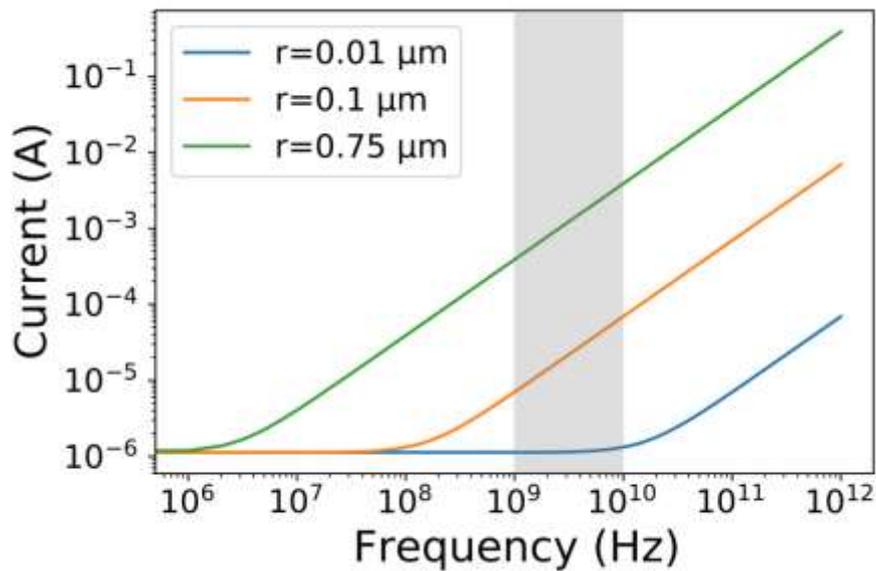

*Figure 4: Plot of current vs. frequency for an MTJ modelled as a parallel RC circuit*

Our findings will be applicable to various MTJ-related applications and RF measurement schemes. The large displacement current we have observed holds significant potential for inducing magnetization dynamics for device applications, such as SOT-induced magnetization switching and the propagation of Oersted-field-driven spin waves. Notably, this displacement does not cause a charge current to flow across the MgO barrier, thus posing only a low risk for barrier breakdown in MTJ devices. However, while these advantages are evident, the displacement current may not be desirable in certain applications, such as high-frequency TMR reading. To address this concern, reducing the capacitance effect can be achieved by shrinking the size of the MTJ pillar (i.e., the area of the parallel plate capacitor). To explore this relationship, we conducted calculations to determine the net current flow for MTJs with radii of 100 and 10 nm under the same voltage conditions as those used in our experiments, as depicted in Fig. 4. For sub-100 nm MTJs, the displacement current in the mA range is expected to be beyond the 100 GHz range.

Finally, our finding also impacts RF techniques such as spin torque [18] or voltage-FMR [33] that are often measured at GHz frequencies, where MTJs of a similar size to those in our experiments are commonly employed. Here, the current flow is usually calculated by assuming MgO as a resistive barrier. However, we conclude that it is important to reevaluate this assumption by considering the MTJ as an RC circuit, where the current flow can be significantly higher. Given that this displacement current does not pass through the MgO barrier, STT might not be present, while the influence of SOT or Oersted field becomes a defining factor in magnetization dynamics, as unequivocally demonstrated in our experiments.

## Acknowledgment

This work is supported by the UKRI-EPSRC Project No. EP/V027808. SC acknowledges the allocation of user time for TR-MOKE measurements in the Exeter Time Resolved Magnetism (EXTREMAG) Facility at the University of Exeter (UKRI-EPSRC grant refs. EP/R008809/1 and EP/V054112/1). WS acknowledges grant no. 2021/40/Q/ST5/00209 from the National Science Centre, Poland. JM acknowledges the program "Excellence initiative research university" for the AGH University of Krakow. FC and LEH acknowledge the Spanish MICIU/AEI/10.13039/501100011033 and the ERDF/EU (Project No. PID2021-122511OB-I00 and "Maria de Maeztu" Units of Excellence Programme No. CEX2020-001038-M). We acknowledge H.Kubota (National Institute of Advanced Industrial Science and Technology, Tsukuba, Japan) for fruitful discussion.

# Supplemental Material

# Magnetization dynamics driven by displacement currents across a magnetic tunnel junction

## Contents

## I. TMR study as a function of MTJ annealing conditions

Using the fabrication procedure explained in the main text, we prepared more than 100 MTJ devices on the same substrate. Then, using an MTJ device similar to the one shown in Fig. 1 in the main text, the TMR was measured as a function of the in-plane field ($B_y$) before annealing, after annealing for 1 hour at 300°C and at 400°C, and the results are shown in Fig. S1. Two important effects of annealing were observed here: 1) the TMR increased, and 2) the magnetic field required to saturate the magnetization in the in-plane direction was increased. These observations confirm that the annealing improved the spin tunneling through the MgO layer, usually attributed to the crystallization of the MgO layer, and increasing the interfacial out-of-plane anisotropy of the bottom Co ferromagnetic layer of the MTJ device [1].

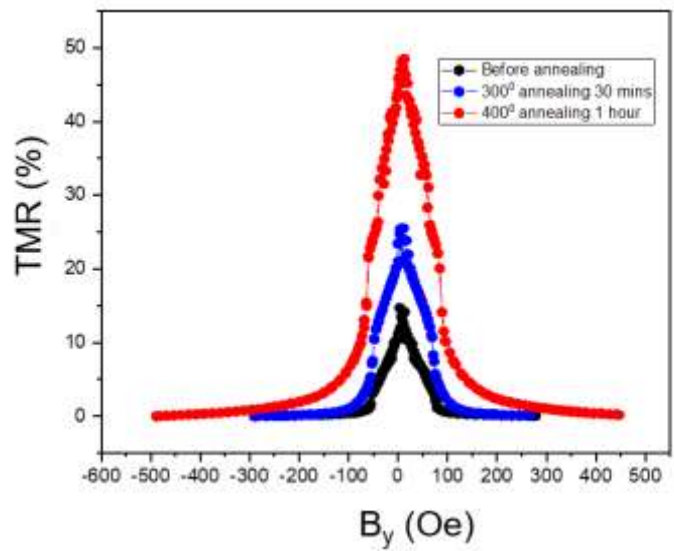

*Figure S5*: TMR vs $B_y$ measurement on an MTJ before annealing (black), after annealing at 300°C (blue), and 400°C (red), respectively.

## II. TR-MOKE measurement of the ferromagnetic resonance

A phase-resolved ferromagnetic resonance (FMR) spectrum was acquired to confirm that measurements with a microwave waveform with frequency of 2 GHz were performed at the FMR resonance magnetic field. The laser spot was positioned on the W/CoFeB microwire near the MTJ ~1 μm away from the top electrode. The TR-MOKE signal was acquired at a fixed time delay of 1.48 ns while the magnetic field was swept. The resulting spectrum is shown in Figure S2 and corresponds to the out-of-plane component of the dynamic magnetization at a time delay close to an antinode of the corresponding TR-MOKE time trace, e.g. 3(e) of main article. The slight bipolar character of the spectrum is due to a mixture of the absorptive and dispersive resonance spectra since the time delay of the measurement (corresponding to the phase of precession) was slightly away from the precession antinode of the TR scan. From the spectrum, an FMR resonance field of 75 mT was identified and used in TR-MOKE measurements with both impulse and microwave waveforms.

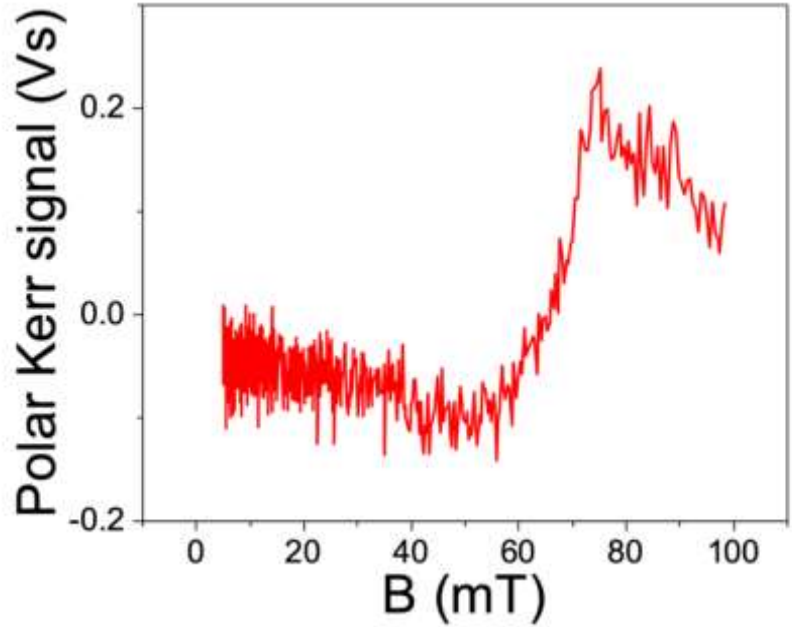

***Figure S2****: The plot of the TR-MOKE signal as a function of the applied magnetic field along y. The laser was focused on the W/CoFeB microwire in regions on the right side of the top electrode. It shows a maximum at ~75 mT, corresponding to the ferromagnetic resonance.*

### III. Calculation of the Oersted field due to a top electrode current

The current through the top electrode can generate an Oersted field on both sides, which was calculated using $B = \frac{\mu_0 I}{2\pi R}$, where $R$ is the distance from the center of the top electrode (Figure S3).

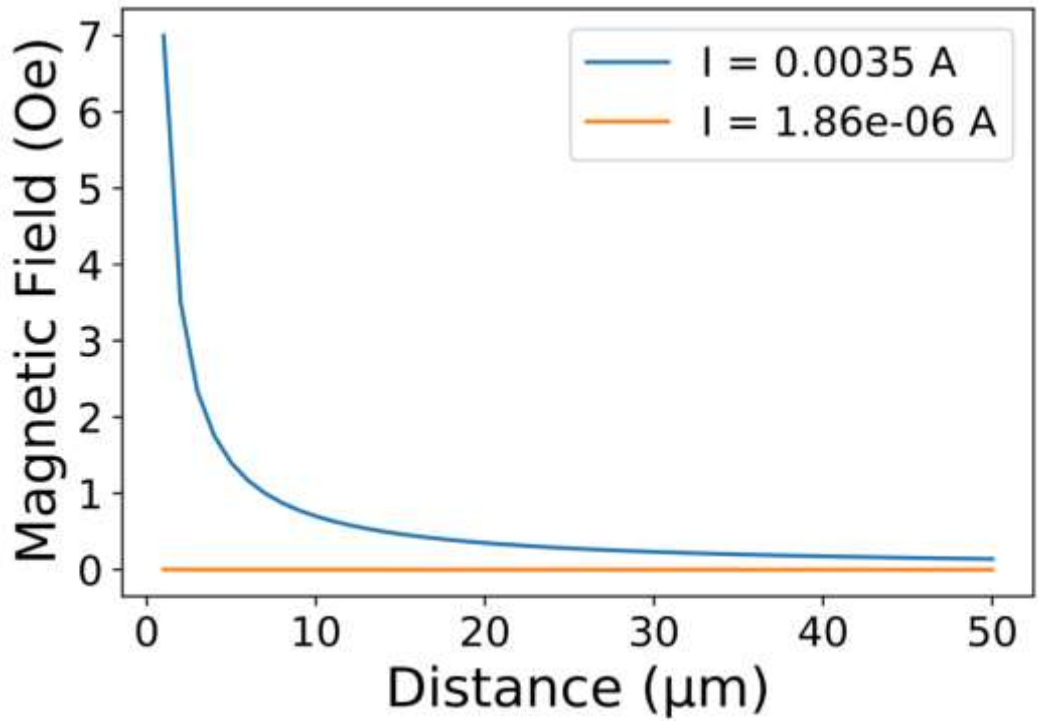

***Figure S3****: Magnetic field as a function of distance from the center of the top electrode for currents of 3.5 mA and 1.86 µA, respectively.*

# IV. SOT and Oersted field-induced magnetization dynamics in the bottom electrode

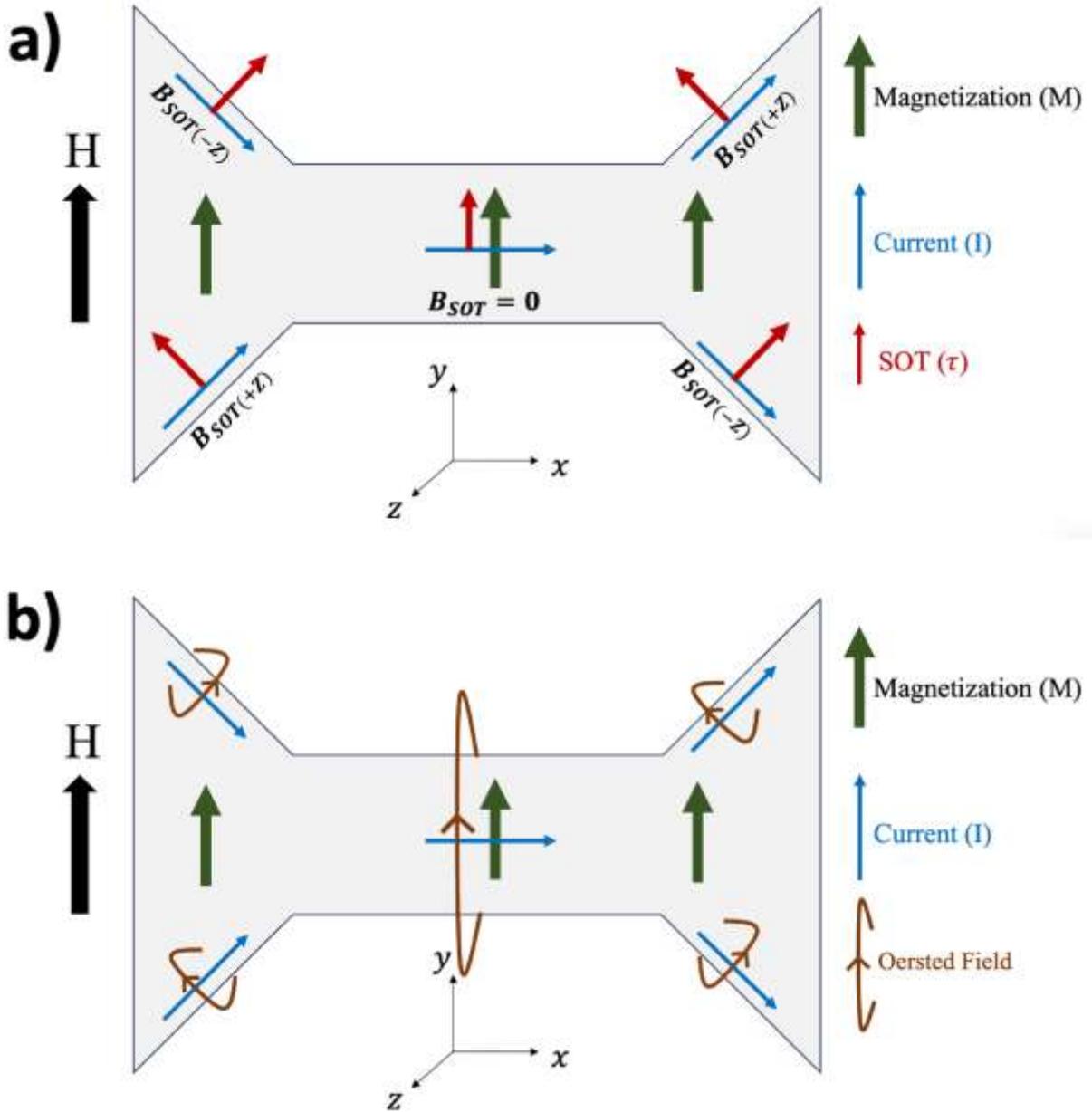

*Figure S4*: Schematic diagrams of magnetization dynamics induced by **(a)** SOT (red arrows) on the magnetization (green) and SOT (red) at different regions of the bottom W/CoFeB electrode. The resulting SOT effective field directions (BSOT) in each direction also marked; **(b)** Oersted field component acting perpendicularly on the magnetization (green) leading to magnetization precession along z, similar to the SOT case explained in panel (a).

The SOT induced magnetization dynamics in the bottom electrode is schematically shown in Figure S4a. The width of the middle region of the W(5 nm)/CoFeB(1.3 nm) wire is ~8 μm. Therefore, a current in above 1 mA is required to reach a current density of $>10^{10}$ mA$^{-2}$, which is crucial for achieving SOT-induced magnetization dynamics. However, the magnetization dynamics occurs when the SOT damping-like torque ($\vec{\tau}$) acts perpendicularly to the magnetization ($\vec{M}$) direction. As shown in Figure S4, the external magnetic field – and thus the magnetization – is fixed along the $y$-axis. In the middle region of the W(5nm)/CoFeB (1.3 nm) bottom electrode, $\vec{\tau}$ is parallel to $\vec{M}$. However, when the current diverges into the triangular region, the $\vec{\tau}$ component along the $x$ direction acts on $\vec{M}$ along $y$, giving an effective magnetic field ($B_{SOT}$) along the $z$-direction, resulting in the magnetization tilting along the $z$-direction giving the observed TR-MOKE contrast. The magnetization dynamics is opposite for current flowing along opposite diagonals in the triangular region giving opposite TR-MOKE contrast, as shown in Fig. 2e in the main text.

The action of Oersted field is explained in the main text and it is schematically shown in Figure S4b. It has the same symmetry as the SOT-induced magnetization dynamics.

### V. Capacitance of a parallel plate capacitor of varying area

The capacitance ($C$) of a parallel plate capacitor is given by

$$C = \frac{Q}{|\Delta V|},$$

where $Q$ and $\Delta V$ are the total charge and potential difference of the parallel plate capacitor, respectively. Further, the potential difference is given by

$$\Delta V = -\int_0^d E \cdot dl \ .$$

Next, the electric fields $E_1$ and $E_2$ of two parallel plates with areas $A_1$ and $A_2$ separated by a distance $d$ can be written as

$$E_1 = \frac{\sigma_1}{2\varepsilon_0} = \frac{Q}{2A_1\varepsilon_0},$$

$$E_2 = \frac{\sigma_2}{2\varepsilon_0} = \frac{Q}{2A_2\varepsilon_0},$$

where $\sigma$ is the charge density. Then, the total potential difference can be written as

$$\Delta V = -\int_0^d (E_1 + E_2).dl$$

$$\Delta V = -\frac{Q}{2\varepsilon_0}\left(\frac{1}{A_1}+\frac{1}{A_2}\right)\int_0^d dl$$

In our case, $A_2 \gg A_1$, thus

$$\Delta V = -\frac{Qd}{2A_1\varepsilon_0}$$

and

$$C = \frac{2A_1\varepsilon_0}{d}.$$

## VI. Charging and discharging of the MTJ capacitor

The charging and discharging of the capacitor as a function of time for $V$=930 mV, $C$=0.13pF and $R_{MTJ}$=490 k$\Omega$ was calculated and plotted in Fig. S5.

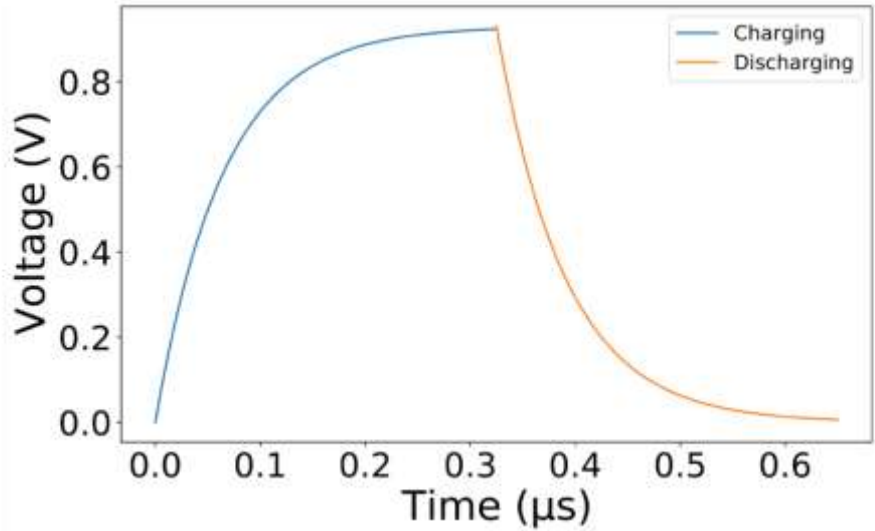

**Figure S5:** *Calculated charging and discharging voltage of the MTJ capacitor as a function of time.*

## VII. Reproducibility

To verify the reproducibility of our results, we conducted the TR-MOKE experiment using a different MTJ device of 10 µm in diameter on the same chip. The MgO interface resistance of this MTJ device was ~14 k$\Omega$, significantly lower than that of the sample discussed in the main text. Following the same procedure explained in the main text, we first performed time-delay TR-MOKE measurements by focusing the laser spot on the right side of the MTJ while applying an input voltage pulse (rise time = 30 ps, pulse width = 70 ps) of 0.5 V and $B_y$ = 75 mT (corresponding result is shown in Fig. S6(a)). Then, spatial scans at time delays of 1.58, 1.83, and 2.08 ns, corresponding to the first +antinode (+AN), node (N), and -antinode (-AN), respectively, were performed as shown in Fig. S6(c). Similar to the results obtained for the device presented in the main text, we observed Oersted field-induced magnetization dynamics which exhibited opposite contrast on the two sides of the top electrode. Compared to the device in the main text, the TR-MOKE contrast was lower here, as expected, due to the smaller MgO interface resistance in this device compared to the device in the main text, resulting in less of a capacitance effect. To further confirm the Oersted field effects, we repeated the TR-MOKE measurements by applying an RF input signal at 2 GHz only through the top electrode, as shown in Fig. S6 (b,d). As expected, a strong TR-MOKE contrast due to Oersted field-induced magnetization dynamics was observed. These results confirm the reproducibility of the displacement current-induced magnetization dynamics effect explained in the main text.

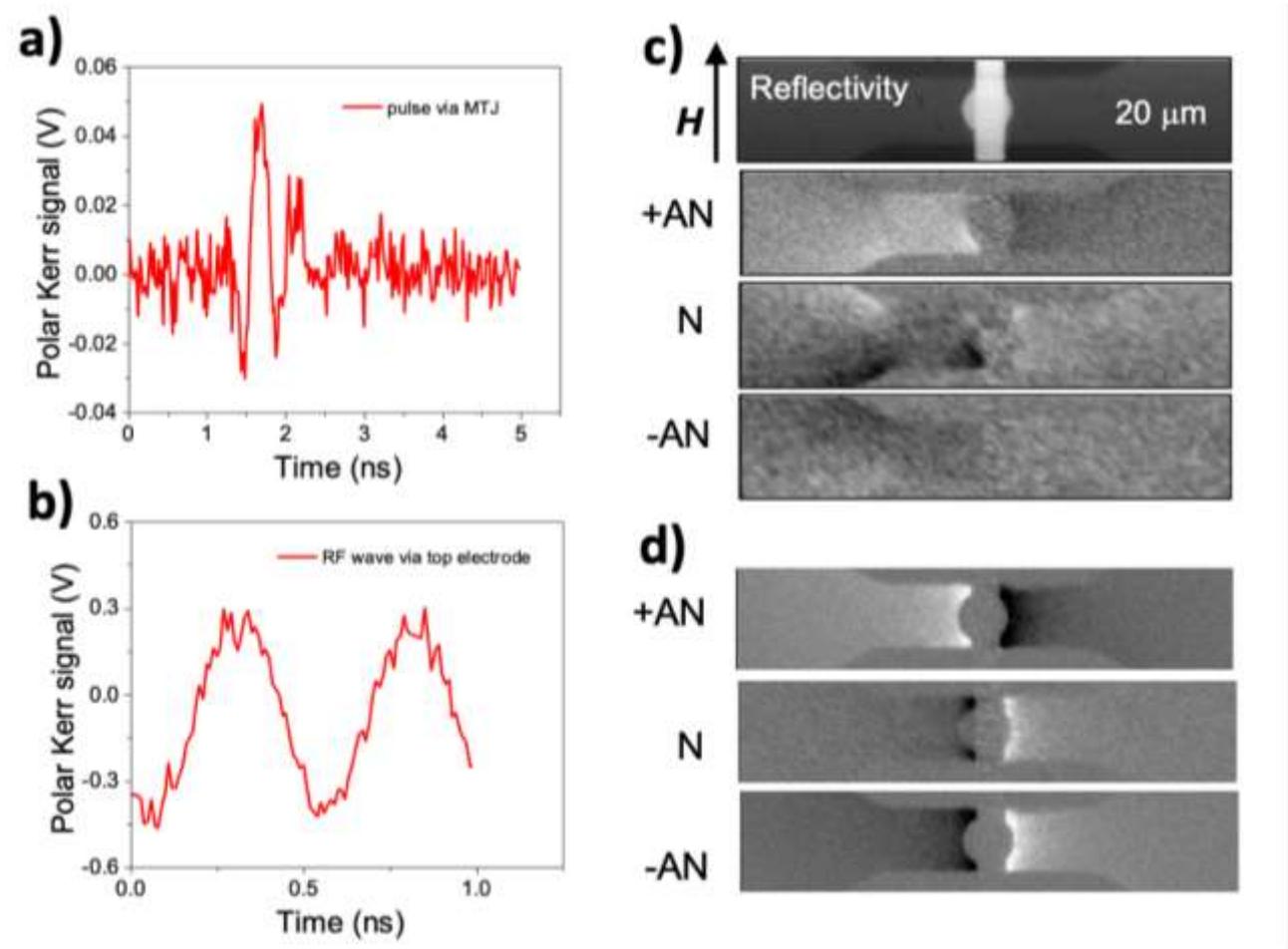

*Figure S6: TR-MOKE experiment results performed using a second MTJ device from the same substrate **(a)** Plot of the Kerr rotation vs. time delay taken after fixing the laser spot at the left side of the top electrode and applying an input voltage pulse=0.5V across the MTJ of resistance = 14 kΩ **(b)** corresponding TR-MOKE scanning images taken at a fixed time delay correspond to the first +antinode (+AN), node (N), and -antinode (-AN) of the plot shown in panel (a). **(c)** The reflectivity image of the real device (above) shows the bottom electrode, MTJ pillar and top electrode. Below, plots of the Kerr rotation vs. time delay are shown, obtained by applying a RF wave of V= 50 mV across top electrode of resistance = 72Ω. **(d)** Scanning TR-MOKE image for the same RF wave conditions as shown in panel (c).*

# VIII. Micromagnetic simulations

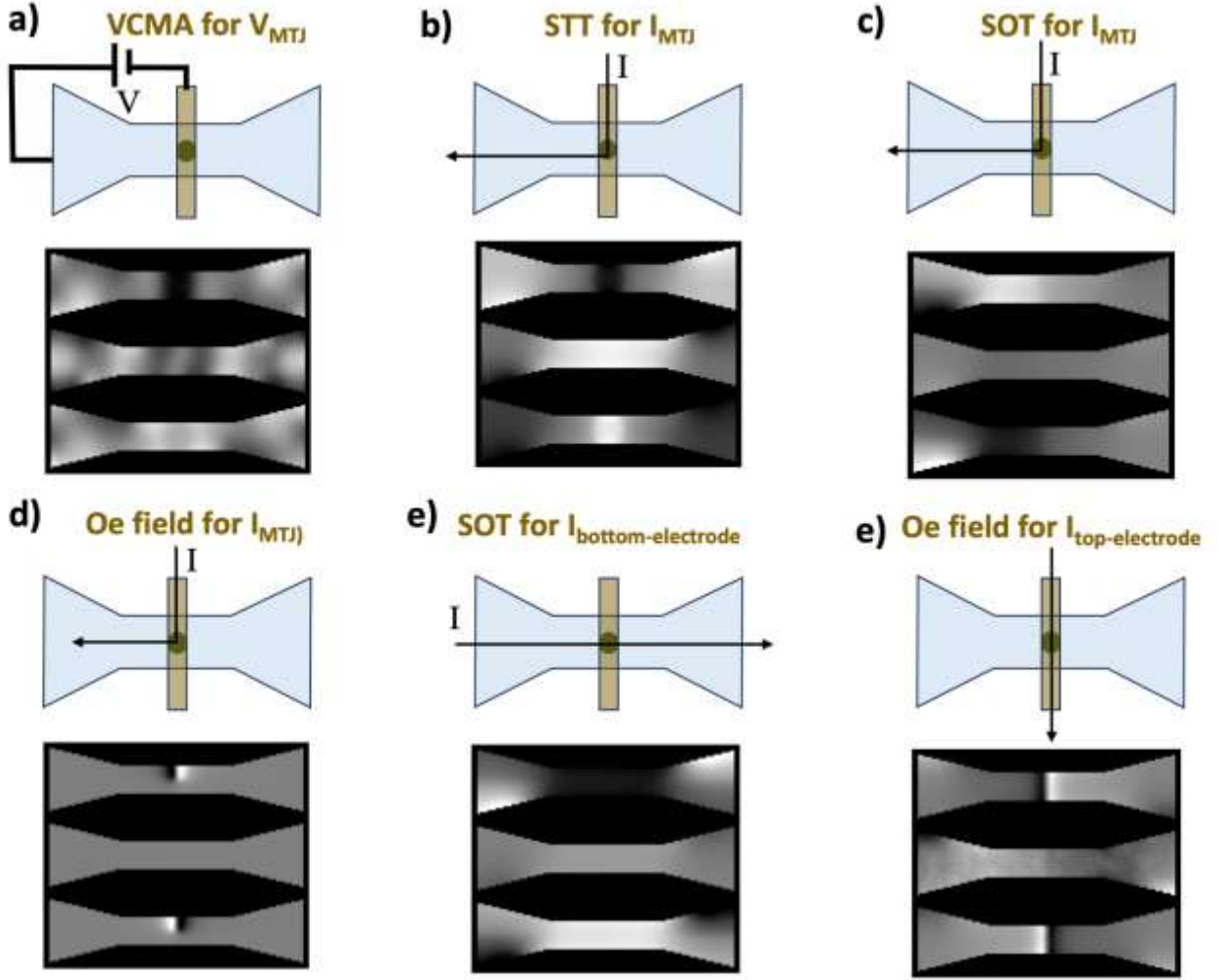

*Figure S7:* Micromagnetic simulations revealing the magnetization dynamics in the bottom electrode for the following scenarios: **(a)** VCMA with voltage across MTJ leads, **(b)** STT with current through MTJ, **(c)** SOT generated in the bottom electrode region where the current flows, **(d)** Oersted field generated from the top electrode region where the current flows, **(e)** SOT in the entire bottom electrode due to the current applied through the bottom electrode only, and **(f)** Oersted field due to the current applied only through the top electrode, which leads to opposite Oersted fields on the two sides of the top electrode. For each case, the top panel shows the current or voltage configuration, and the bottom image three snapshots of the simulated magnetization dynamics at time=maximum, minimum and zero nodes. The black and white contrast corresponds to the magnetization oriented along the +Z and -Z directions, respectively.

To qualitatively explore the possible magnetization dynamics in our device due to VCMA, STT, SOT and an Oersted field, we performed micromagnetic simulations with mumax3 [2]. The different magnetic parameters used for the simulation, such as cell size, exhcnage energy constant ($A$), saturation magnetization ($M_s$) and anisotropy energy density ($K_u$) and magnetization damping ($\alpha$) are shown in Table 1. First, we saturate the magnetization by applying an in-plane field, $B_y$, of 75 mT. For each excitation mechanism, we use an RF signal in the frequency range from 3 to 9 GHz (chosen to maximize the reponse), and take a snapshot at the maximum, minimum and zero nodes as shown in Fig. S7. In case of VCMA we asummed the energy change of 50 fJ/Vm [1]. The experimentally applied magentic field of 75 mT prevents VCMA induced dynamics, therefore, in the VCMA simulation case, we lower the By field to 1 mT, and increase the anisotropy to better capture the excitation via the RF signal.

In the STT and VCMA simulations (Fig.S7(a,b)), we excite the system exclusively in the center, where the circular MTJ pillar meets the FM layer. Similarly, for the case (c), where the system is excited with the current flowing through the top electrode and then through the pillar, we model SOT excitation only in the left half of the geometry.

| Parameter | Value | Unit |
| --- | --- | --- |
| Cell size | 2 | nm |
| $A$ | 13 | pJ/m |
| $M_s$ | 1.27 | MA/m |
| $K_u$ | 0.2 (VCMA: 0.8 ± 0.2)* | MJ/m$^3$ |
| $\alpha$ | 0.01 | - |

**Table 1:** *Summary of the magnetic parameters used in our micromagnetic simulations. $A$, $M_s$, $K_u$, and $\alpha$ are the exchange constant, saturation magnetization, anisotropy energy density and damping constant, respectively. *Note that $K_u$ is increased for the VCMA case to better excite the system.*

The micromagnetic simulation results shown in Figs.S7(a-d) reproduce the magnetization dynamics illustrated in Figs.3(a-d) of the main text well. This further confirms that the magnetization dynamics for the input excitation through the MTJ, shown in Fig. 2(c), is due to the Oersted field. The simulation results shown in Figs. S7(e,f) support our experimental results shown in Figs. 2(d.e) in the main text.

IX. **Substrate RF current effects**

Recent studies report that RF magnetotransport measurements can be affected by the capacitive current flowing through the SiO$_2$/Si substrate [3,4], depending on the conductivity of Si substrate. In our device, the resistivity of Si can be expected to be between 0.1 to 100 Ω.cm. Therefore, the substrate may shunt the RF current flowing through the top MgO capacitor circuit. Even though our experiment confirms that there is a large current flowing through the top electrode above MgO, leading to the measured TR-MOKE contrast due to the Oersted-field induced magnetization dynamics, we performed calculations of the substrate RF current. The equivalent circuit is shown in Fig. S8(a). The calculation of the displacement current through the entire circuit, through the substrate, and through the MTJ for resistivities of 0.1 and 100 Ω.cm are shown in Fig. S8(b) and S8(c), respectively.

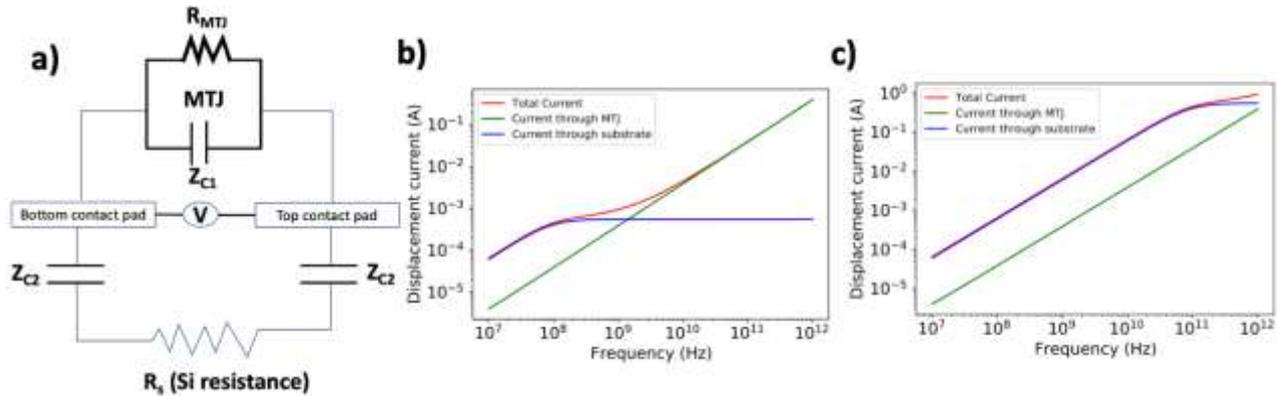

***Figure S8:*** *(**a**) The equivalent electrical circuit consists of the MTJ RC circuit which is parallel to the substrate RC circuit. $Z_{C1}$ is the impedances of MTJ capacitor, while $Z_{C1}$ is the impedance of the capacitor formed between top/bottom contacts and the Si substrate, separated by 500 nm of the dielectric $SiO_2$. The calculated RF current through MTJ capacitor, substrate capacitor and total current are plotted as a function of frequency for Si resistivity (**b**) 0.1, and (**c**) 100 Ω.cm.*